\documentclass[twocolumn,superscriptaddress,amsmath,amssymb]{revtex4-2}
\usepackage{color}
\usepackage{graphicx}
\usepackage{subfigure}
\usepackage{booktabs}
\usepackage{dcolumn}
\usepackage{bm}
\usepackage{amsmath}
\usepackage{verbatim}
\usepackage{lineno}
\usepackage{xcolor} 
\usepackage{float} 
\usepackage{placeins} 
\usepackage{cancel} 

\newcommand{\beq}{\begin{eqnarray}}
\newcommand{\eeq}{\end{eqnarray}}

\newcommand{\red}[1]{\textcolor{red}{#1}}

\newcommand\HG[1]{\textcolor{violet}{ [HG:\,#1]}}

\usepackage[normalem]{ulem}

\begin{document}

\title{Liquid and solid layers in a thermal deep learning machine}


\author{Gang Huang}
\affiliation{Department of Physics, Chengdu University of Technology, Chengdu 610059, China}
\affiliation{Institute of Theoretical Physics, Chinese Academy of Sciences, Beijing 100190, China}

\author{Lai Shun Chan}
\affiliation{Department of Physics, City University of Hong Kong, Hong Kong Special Administrative Region of China, People's Republic of China}

\author{Hajime Yoshino}
\affiliation{D3 Center, Osaka University, Toyonaka, Osaka 560-0043, Japan}
\affiliation{Graduate School of Science, Osaka University, Toyonaka, Osaka 560-0043, Japan}

\author{Ge Zhang}
\email{gzhang37@cityu.edu.hk}
\affiliation{Department of Physics, City University of Hong Kong, Hong Kong Special Administrative Region of China, People's Republic of China}

\author{Yuliang Jin}
\email{yuliangjin@mail.itp.ac.cn}
\affiliation{Institute of Theoretical Physics, Chinese Academy of Sciences, Beijing 100190, China}
\affiliation{School of Physical Sciences, University of Chinese Academy of Sciences, Beijing 100049, China}
\affiliation{Wenzhou Institute, University of Chinese Academy of Sciences, Wenzhou, Zhejiang 325000, China}

\date{\today}

\begin{abstract}
Based on deep neural networks (DNNs), deep learning has been successfully applied to many problems, but
its mechanism is still not well understood -- especially the reason why over-parametrized DNNs
can generalize.
A recent statistical mechanics theory on supervised learning by a prototypical multi-layer perceptron (MLP) on some artificial learning scenarios predicts that adjustable parameters of over-parametrized MLPs become strongly constrained by the training data close to the input/output boundaries, while the parameters in the center remain largely free, giving rise to a solid-liquid-solid structure.  
Here we establish this picture, through numerical experiments on benchmark real-world data
using a thermal deep learning machine that explores the phase space of the synaptic weights and neurons.
The supervised training is implemented by a GPU-accelerated molecular dynamics algorithm, which operates at very low temperatures, and the trained machine exhibits good generalization ability in the test. 
Global and layer-specific dynamics, with complex non-equilibrium aging behavior,  are characterized by time-dependent auto-correlation and replica-correlation functions.
Our analyses reveal that the design space of the parameters in the liquid and solid layers are respectively structureless and hierarchical. Our main results are summarized by a data storage ratio - network depth phase diagram with liquid and solid phases.
The proposed thermal machine, which is a physical model with a well-defined Hamiltonian, 
that reduces to MLP in the zero-temperature limit, can serve as a starting point for physically interpretable deep learning.

\end{abstract}

\maketitle

\begin{figure}[!htbp]
  \centering
  \includegraphics[width=\linewidth]
  {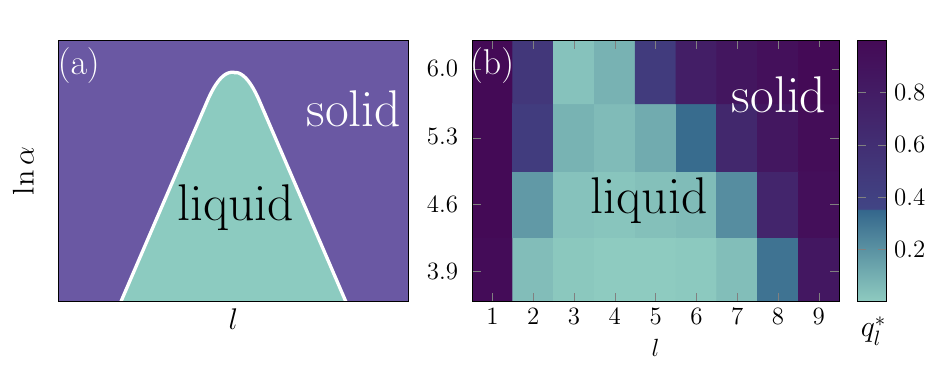}
  \caption{
  {\bf Liquid-solid phase diagram of deep learning.}
  (a) Schematic phase diagram for the state of network parameters in the $l$-th layer with a given capacity ratio $\alpha=M/N$, drawn according to the theoretical prediction Eq.~(\ref{eq:xig}) 
  in $N \to \infty$ limit~\cite{Yoshino2020, yoshino2023spatially}. 
  (b) Numerical phase diagram obtained by MD learning simulations, color coded according to the spin overlap between replicas, $q_l^* \equiv q_l(t=t^* = 5\times 10^5, t_{\rm w} = 5 \times 10^4)$. 
  Here $L=10, M=2000$ are fixed, and $N$ is varied. 
  The boundary between liquid (green) and glass (blue)  phases is estimated approximately at $q_l^* \approx 1/e$.}
  \label{fig:phase_diagram}
\end{figure}

{\bf Introduction.}
In deep learning, a fundamental problem is to understand how the mapping between the input data ${\bf S}_0$ and the output data ${\bf S}_L$ is established through intermediate hidden layers of neurons, denoted by {\it internal representations} ${\bf S}_l$~\cite{Monasson1995}, where $l=1 \ldots L-1$ and $L$ is the number of layers in the DNN. 
Noticeable progress has been made 
within the framework of statistical mechanics,  based on the intrinsic connection between neural networks and spin glass problems ~\cite{Hopfield1982, engel2001statistical, Carleo2019, Nishimori2001, Gardner1987, Gardner1988,  baldassi2016unreasonable, choromanska2015loss, ballard2017energy, Baity-Jesi2018, franz2019jamming, geiger2019jamming,  Yoshino2020, yoshino2023spatially, huang2024energy, winter2024glassy}. 
The statistical mechanics theory of supervised learning of single perceptrons by Gardner
\cite{Gardner1987, Gardner1988} has been extended to MLPs by Yoshino~\cite{Yoshino2020,  yoshino2023spatially}.
Such a theory considers ensembles of machines specified by 
realizations of synaptic weights (connection weights between neurons) ${\bf J}_l$ 
as well as the neurons (internal representations)~\cite{Monasson1995} in the hidden layers  ${\bf S}_l$.
The phase space volume of the machines, called {\it Gardner's volume}, is the number of machines
that meet all the constraints imposed by the training data. Such constraints are specified by the boundary conditions: 
${\bf S}_0$ and ${\bf S}_L$ are quenched boundaries that may be  (i) generated simply randomly 
(random scenario: a random constraint satisfaction problem) 
or (ii) provided through a randomly generated teacher-machine (teacher-student scenario: a statistical inference problem). The replica theory for the MLP finds that typical machines in Gardner's volume
exhibit a solid-liquid-solid structure with the hidden layers near boundaries in the solid states
and the middle layers in the liquid states (see Fig.~\ref{fig:phase_diagram}a). 
The solid state corresponds to a glass in the (i) random scenario, and a crystal associated with the teacher machine in the (ii) teacher-student scenario.
Such a structure could also balance representation and generalization, via respectively the vast microscopic states in liquid layers and the strong correlations to the input/output boundaries in solid layers.

An interesting implication of the theory, which is based on the internal representation,
is that one can replace the conventional machine learning algorithms based on back-propagation
by physically more sensible ones, called {\it thermal deep learning machines (TDLM)}.  The conventional algorithms 
try to minimize a loss function that depends on the actual labels ${\bf S}_L$, and the predicted outputs ${\bf S}'_L$ 
generated by a sequence of highly convoluted linear and non-linear operations 
on ${\bf S}_{0}$. In particular, such a conventional loss function does not explicitly depend on the internal representation. 
In contrast, the TDLM has a 
well-defined Hamiltonian consisting of local interactions among the internal representation $\{ {\bf S}_l \}$  and $\{ {\bf J}_l \}$, with  the training data imposed as fixed boundary conditions ${\bf S}_{0}$ and ${\bf S}_{L}$. With the given Hamiltonian, the TDLM can be simulated at any temperature. 
Under the framework of TDLMs, the learning problem can be conveniently tackled using theoretical and numerical tools developed in statistical physics.

In physics, there are numerous examples of boundary-induced phenomena, including
liquid-gas wetting~\cite{nelson2004statistical}, melting dynamics~\cite{krzakala2011meltingA, krzakala2011meltingB}, the point-to-set correlation length in supercooled liquids~\cite{biroli2008thermodynamic, kob2012non, hocky2012growing, ikeda2015one}, and vapor-deposited ultrastable glasses~\cite{singh2013ultrastable, berthier2017origin}.
Deep learning provides another interesting playground  where 
breaking of the translational symmetry imposed by the boundaries can bring non-trivial consequences.

Although the picture provided by the replica theory for MLP is interesting, the current theory considers completely random, structureless input data~\cite{Yoshino2020, yoshino2023spatially}, 
which is very far from reality. Also, as a static theory, the replica theory does not address training dynamics. 
In order to examine the theoretical picture in deep learning of real-world data, one has 
to perform numerical experiments. This is the motivation of the current study. 

Our simulations establish the coexistence of liquid and solid layers in learning dynamics,  summarized by a numerical phase diagram (see Fig.~\ref{fig:phase_diagram}b).  For MNIST benchmark data~\cite{LeCun, Qiao2007}, the TDLM trained by a GPU-accelerated molecular dynamics (MD) algorithm, has good generalization with a test accuracy 
up to $\sim 98 \%$. The training procedure contains three dynamical regimes, with strong aging effects appearing in the intermediate time regime. Our analyses based on time-correlation functions and replica overlap order parameters show distinct dynamic and thermodynamic features of the network layers in the center and near the boundaries. The states in the central layers are liquid-like, characterized by rapid relaxation dynamics and nearly zero overlaps between differently trained machines. In contrast, the states in the near-boundary layers are solid-like, characterized by slow dynamics and a hierarchical organization of the overlap order parameters. 
 
{\bf A thermal machine for supervised deep learning.}
The task is to classify handwritten digits taken from the MNIST database~\cite{LeCun, Qiao2007}, with $M$ training images and 400 test images (see Appendix A). 
The TDLM employs the internal representation so that its Hamiltonian (see Appendix B) involves $\{ {\bf S}_l \}$ and $\{ {\bf J}_l \}$ as free dynamical variables. 
The canonical ensemble of the model becomes identical to the ensemble of
fully connected MLPs in the zero temperature limit.
We consider a rectangular network with width (the number of perceptrons in a hidden layer) $N$ and depth (number of layers) $L$ (see  Appendix B). 
The {\it storage ratio} $\alpha = M/N$ quantifies the average degree of constraints per free variable in a layer. 
The training is accomplished by relaxing the total energy $E$ (the expectation value of the Hamiltonian) using MD simulations at 
a very low temperature $T = 10^{-5}$ (see SI Sec.~S1 for other $T$), starting from random configurations (see Appendix C). 

This model is studied theoretically by Yoshino in Ref.~\cite{Yoshino2020, yoshino2023spatially}
in the zero temperature limit under the assumption that input data are completely random, 
and the correlations caused by interaction loops can be neglected 
(which is justified by considering the dense limit $ N \gg c \gg 1$ with $c$ being
the connectivity of the inter-layer coupling.) 
The key prediction is that  the {\it penetration depth} $\xi_{\rm s}$ of solid layers 
(either glass or crystal) near the boundaries   linearly depends on $\ln \alpha$,
\beq
\xi_{\rm s} \sim \ln \alpha,
\label{eq:xig}
\eeq
according to which a schematic phase diagram is plotted in Fig.~\ref{fig:phase_diagram}a.

\begin{figure*}[!htbp]
     \includegraphics[width=0.9\linewidth]{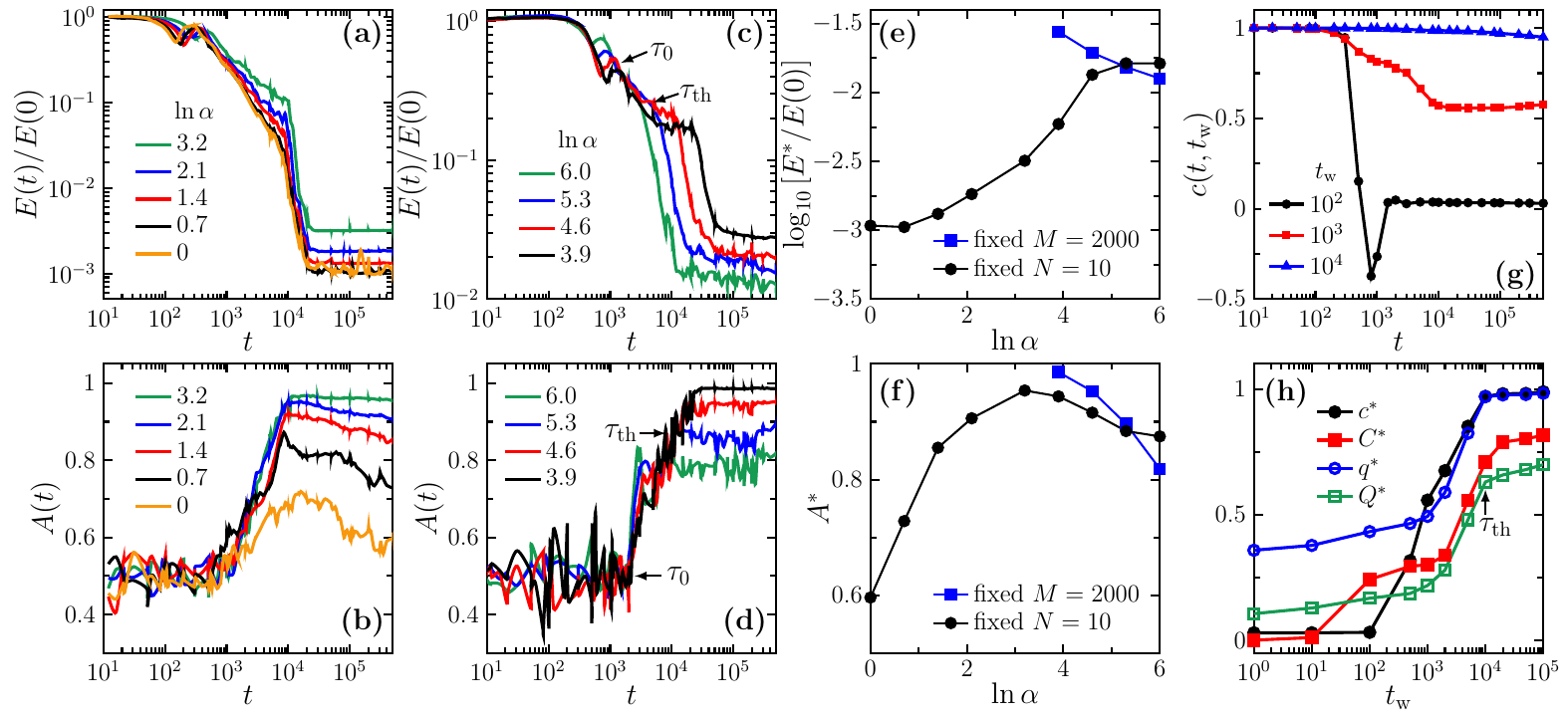}
    \caption{
    {\bf Global training dynamics and test accuracy. }
   Time evolution of the (a,c) {reduced potential energy $E(t)/E(0)$} and (b,d) test accuracy $A(t)$, for different $\ln \alpha = \ln(M/N)$.
   In (a,b), $M=2000$ is fixed and in (c,d) $N=10$ is fixed (see Table S1). 
   (e,f) $E^*=E(t^*)$ and $A^*=A(t^*)$, where $t^*=5 \times 10^5$ is the maximum simulation time. 
 (g) Global  auto-correlation function of spins $c(t, t_{\rm w})$.
   (h) The values of different correlation functions at $t^*$ are plotted as functions of $t_{\rm w}$.    In (g,h),  $\ln \alpha = 5.3$ and $M=2000$, for which $\tau_0 \approx 10^3$ and $\tau_{\rm th} \approx 10^4$ (marks). 
 }
    \label{fig:energy_accuracy}
\end{figure*}

{\bf Global training dynamics and generalization tests.}
Figure~\ref{fig:energy_accuracy}(a-d) shows the time evolution of the reduced total potential energy $E(t)/E(0)$
during the MD training, and the generalization ability $A(t)$ as the proportion of correctly labeled images in the test dataset.
To obtain $A$, we switch back to the standard feed-forward DNN (see Appendix D), using the instantaneous $\{ {\bf J}_l (t) \}$ generated by the TDLM during training.
 Note that in the test, $\{ {\bf S}_l \}$ are not free variables 
 but determined by the input test data, weights $\{ {\bf J}_l \}$, and the activation function.  
 The training energy $E^*$ and test accuracy $A^*$ at the largest simulation time, $t^*=5 \times 10^5$, are plotted in Fig.~\ref{fig:energy_accuracy}(e,f). 
 Non-monotonic $A^*(\alpha)$ is observed, with the maximum separating over-parameterized ($\ln \alpha < 3.2$) and over-constrained  ($\ln \alpha > 3.2$) regimes. 
The initial increase of $A^*(\alpha)$ with the number of constraints $\alpha$ agrees with the
general expectation that the generalization ability should improve by learning more data.
On the other hand, the decrease of  $A^*(\alpha)$ with too strong constraints implies frustration effects. On the theoretical side, it has been observed that the idealized (Bayes optimal) teach-student scenario develops a crystal phase \cite{yoshino2023spatially}, while replica symmetry breaking takes place away from such an idealized situation \cite{Yoshino2020}.
 
According to Fig.~\ref{fig:energy_accuracy}(a-d), glassy-like two-step learning dynamics can be identified. (i) For $t< \tau_0$, where $\tau_0$  is a microscopic time scale,  $A(t) \approx 0.5$, is nearly constant, while $E(t)$ decreases due to random thermal fluctuations. 
(ii) In the intermediate regime, $\tau_0 < t <\tau_{\rm th}$, where $\tau_{\rm th}$ is a thermalization time, $E(t)$ decreases and $A(t)$ increases logarithmically $\sim \ln t$. This non-equilibrium aging region is responsible for learning as the test accuracy is significantly improved. The logarithmic aging behavior, $E(t) \sim - \ln t$, is typical in spin glass models such as the random energy model~\cite{li2024simplest}. The logarithmic decay of the loss function is also observed in stochastic gradient descent training of standard MLMs~\cite{Baity-Jesi2018}. 
(iii) After $\tau_{\rm th}$, $E(t)$ rapidly decays to a near-zero value, and $A(t)$ does not increase anymore. 

To further characterize the non-equilibrium dynamics, we measure four types of time-dependent correlation functions  (see Appendix E for definitions).
Figure~\ref{fig:energy_accuracy}(g) shows $t_{\rm w}$-dependence of the spin auto-correlation function $c(t,t_{\rm w})$ for intermediate $t_{\rm w}$ with $\ln \alpha = 5.3$, which confirms the existence of aging effects (see SI Sec.~S2 for additional data). 
In Fig.~\ref{fig:energy_accuracy}(h), we plot the values of correlation functions ($c^*, C^*, q^*, Q^*$) at a large time $t^*=5 \times 10^5 > \tau_{\rm eq}$ (e.g., $c^* \equiv c(t=t^*, t_{\rm w})$), as functions of $t_{\rm w}$. The intermediate $\ln t_{\rm w}$ behavior is robustly observed. Moreover, these values become nearly $t_{\rm w}$-independent after $\tau_{\rm th} \approx 10^4$, which seems to suggest that the system is overall thermalized.

{\bf Layer-dependent training dynamics.}
The layer-dependent training dynamics are revealed by  the layer specific correlation functions (see Appendix E for definitions). 
The results in Fig.~\ref{fig:layer_dynamics} are obtained with $\ln \alpha = 5.3$.
Figure~\ref{fig:layer_dynamics}a shows $c_l(t,t_{\rm w})$ for the auto-correlations of spins in the $l$-th layer, for $l=5$ and $9$. Obviously, the 5-th layer in the center de-correlates much faster than the 9-th layer near the output boundary, suggesting respectively liquid-like and solid-like behavior. 
The relaxation time $\tau_{l}^c$, defined by $c_l(t=\tau_{l}^c, t_{\rm w}) =  1/e$, is plotted in Fig.~\ref{fig:layer_dynamics}b as functions of $t_{\rm w}$.
For the  central layers ($l=3-5$),  $\tau_{l}^c$ reaches a plateau for $t_{\rm w} > \tau_{\rm th} \approx 10^4$. In other layers, $c_l(t,t_{\rm w})$ decays too slow to reach the threshold $1/e$ up to the maximum time $t=5\times 10^5$, when  $t_{\rm w}$ is large (in this case, $\tau_{l}^c$ cannot be estimated). In other words, the  $\tau_{\rm th}$ in global dynamics (Fig.~\ref{fig:energy_accuracy}) corresponds to the thermalization time of central liquid layers, while the near-boundary solid layers are still out-of-equilibrium at this time scale.

Similar layer-dependent properties are captured by the layer-specific replica-correlation function $q_l(t, t_{\rm w})$ of spins. 
With a fixed  $t_{\rm w} = 5 \times 10^4$, 
$q_l(t, t_{\rm w})$
decays monotonically with $t$ (see Fig.~\ref{fig:layer_dynamics}c). However, the dependence of $q_l(t, t_{\rm w})$  on $l$ is non-monotonic. As a practical definition, the layers with $q^*_l \equiv q_l(t=t^* = 5\times 10^5, t_{\rm w}=5\times 10^4)< 1/e$  are called  liquid layers, and otherwise solid layers. 
The liquid and solid depths are $\xi_{\rm l}$ and $\xi_{\rm s} = L-\xi_{\rm l}$ respectively (see Fig.~\ref{fig:layer_dynamics}d). 
We use this method to quantitatively determine liquid and solid layers in the network. 
The results with varied $\ln \alpha$ are summarized by the numerical phase diagram Fig.~\ref{fig:phase_diagram}b. The theoretical linear relation Eq.~(\ref{eq:xig}) of the phase boundary is consistent with our numerical data (see SI Secs. S3 and S4 for additional data). 

It is interesting to observe a similar solid-liquid-solid network structure of the internal representations $\{ S_l\}$ during testing (Fig.~\ref{fig:layer_dynamics}d), where $\{ {\bf J}_{l}\}$ are taken from MD trained configurations at the same $t_{\rm w}$ and $t$. 
This structure has significant meaning in the balance between representation and generalization. 
While the correlations between replicated machines are small in the central liquid layers, providing a powerful representation ability, they do not completely vanish. Moving closer to the output boundary, correlations are recovered, giving rise to the generalization ability.

\begin{figure}[!htbp]
     \includegraphics[width=\linewidth]{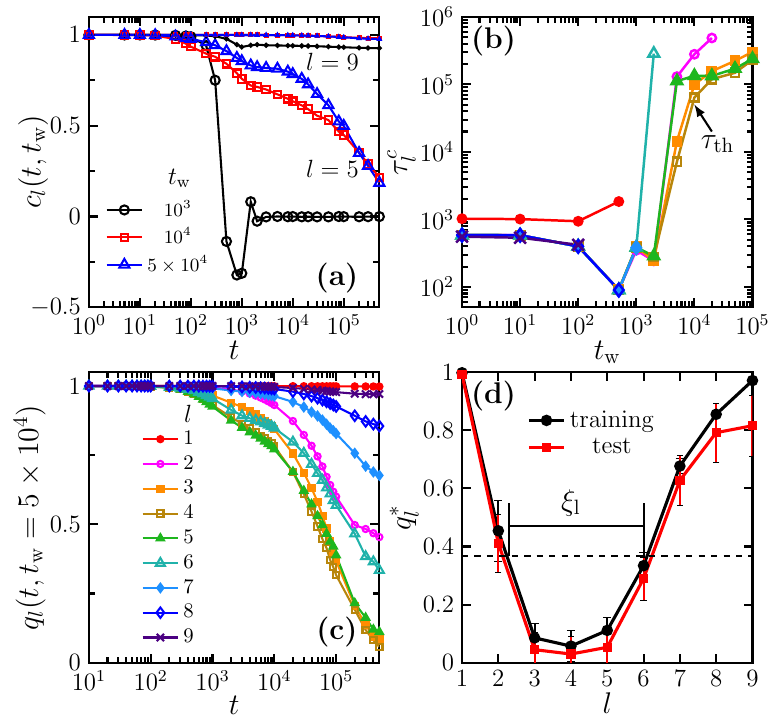}
    \caption{
    {\bf Layer-dependent training dynamics. }
  (a) Spin auto-correlation functions $c_l(t, t_{\rm w})$ in the $l=5$ (open symbols) and $l=9$ (filled symbols) layers, for $t_{\rm w} = 10^3$ (black), $10^4$ (red), $5 \times 10^4$ (blue). 
  (b) Spin correlation time $\tau_l^{\rm c}$ as a function of $t_{\rm w}$, with $\tau_{\rm th} \approx 10^4$ indicated, for different $l$ (colors have the same meaning as in (c)). 
  (c) Spin replica-correlation functions $q_l(t, t_{\rm w}=5 \times 10^4)$ in different layers. (d) $q_l^* \equiv q_l(t=5 \times 10^5, t_{\rm w} = 5 \times 10^4)$ as a function  of $l$. For comparison, the spin overlaps of the internal representations generated during testing are also plotted. 
    We fix $\ln \alpha =5.3$ and $M=2000$ in all panels.
  Error bars represent standard errors.   }
    \label{fig:layer_dynamics}
\end{figure} 


 \begin{figure}[!htbp]
  \includegraphics[width=\linewidth]{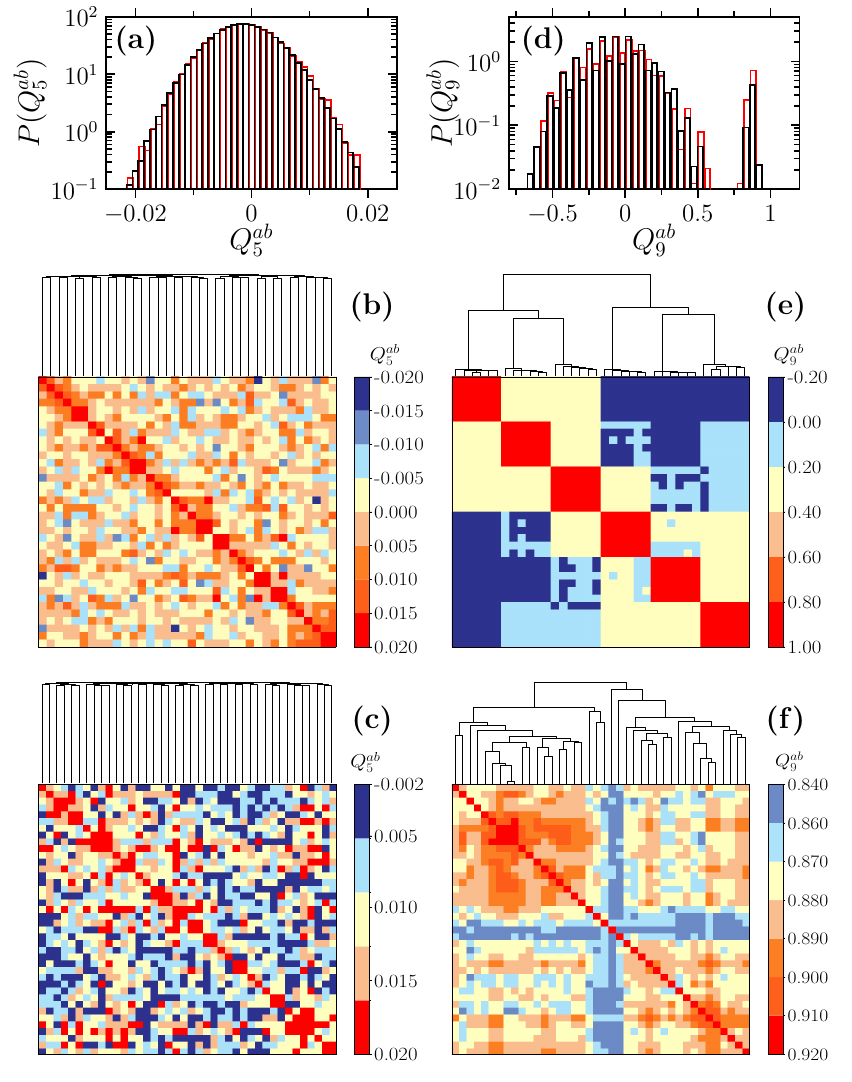}
     \caption{
     {\bf Structure of the design space in typical liquid and solid layers revealed by overlap order parameters.}
     Data in (a-c) and (d-f) are respectively for typical liquid ($l=5$) and solid ($l=9$) layers.
     We set $M=2000$ and  $\ln \alpha = 4.6$.
     (a,d) Probability distribution $P(Q_l^{ab})$, for $(N_{\rm s},N_{\rm r}) = (20, 8)$ (red) and $(38,16)$ (black).
     (b,e) The $36 \times 36$ overlap matrix of $Q_l^{ab}$ obtained with $(N_{\rm s},N_{\rm r}) = (6, 6)$, where the color bar represents the value of the matrix element. 
     (c,f) A close-up of the second sample in (b,e) by increasing the number of replicas from $N_{\rm r}=6$ to $N_{\rm r}=40$ for this sample. 
     }
      \label{fig:overlap_heatmap_dendrogram}
\end{figure} 

{\bf Hierarchical structure of the design space in the solid layers.}
To explore the layer-dependent structure of the design space formed by the micro-states of $\{{\bf J}_{l} \}$, we employ the method of overlap order parameters. The overlap parameter $Q_l^{ab} \sim {\bf J}_{l}^a \cdot {\bf J}_{l}^b$ quantifies the similarity between two configurations $a$ and $b$.
In this study, $Q_l^{ab}$ is estimated from the dynamical data,  $Q_l^{ab} \equiv Q_l^{ab}(t=10^5, t_{\rm w}=10^3)$. 
The following three complementary representations reveal consistently the structural features of the design space (Fig.~\ref{fig:overlap_heatmap_dendrogram}).
(i) The distribution $P(Q_l^{ab})$ computed from $(N_{\rm s} \times N_{\rm r}) \times (N_{\rm s} \times N_{\rm r}) $ pairs of configurations  ($a,b = 1,2, \ldots, N_{\rm s} \times N_{\rm r}$), with  $N_{\rm s}$ independent  random initial configurations at $t=0$ (samples), and $N_{\rm r}$ replicas for each initial configuration replicated at $t_{\rm w}$ (see SI Sec.~S5 for additional data). 
(ii) 
An overlap matrix  whose element is $Q_l^{ab}$. 
The matrix element is sorted using the Complete Linkage agglomerative clustering algorithm~\cite{Hastie2009} (see Appendix F). This method can identify hierarchical groups of configurations with similar overlaps. (iii) Based on the clustered groups, the states are organized into a tree structure. The configurations in the lowest sub-trees have the largest overlaps $Q_l^{ab}$, and configurations with smaller overlaps are grouped in higher-level sub-trees. The procedure is performed recursively in the descending order of $Q_l^{ab}$.

In a typical liquid layer ($l=5$), the states are well-connected and the design space is  structureless.
This property is reflected by a Gaussian distribution  
$p(Q_l^{ab})$ with a zero mean, a random matrix, and a single-level tree (Fig.~\ref{fig:overlap_heatmap_dendrogram}a-c).

The design space structure in a solid layer ($l=9$) is more complex (Fig.~\ref{fig:overlap_heatmap_dendrogram}d-f).  As expected by the theory~\cite{Yoshino2020, yoshino2023spatially}, crystal-like solid layers are essential for generalization, while glassy states may emerge in these solid layers due to frustration caused by a large number of constraints. This expectation is consistent with our numerical results.
The $P(Q_l^{ab})$ in Fig.~\ref{fig:overlap_heatmap_dendrogram}d has two peaks: the right one corresponds to the overlaps between  configurations replicated from the same sample (same solid), and the left one corresponds to those in different samples (different solids).
Multiple solid states appear due to the permutation symmetry in the fully connected DNN. What is more fascinating
is that, within a solid state picked up by a given sample, delicate hierarchical structures can be identified.
We zoom in on one typical sample by increasing the number of replicas $N_{\rm r}$ for the given sample. As shown in Fig.~\ref{fig:overlap_heatmap_dendrogram}f, the diagonal block in the matrix of Fig.~\ref{fig:overlap_heatmap_dendrogram}e actually contains hierarchical sub-blocks, and correspondingly the sub-tree has hierarchical sub-sub-trees. Thus the design space in  a crystal state is not structureless, but instead consists of hierarchical clusters of solutions as suggested by the mean-field  calculations~\cite{Yoshino2020}. A similar picture has been proposed for the phase space of spin and structural glasses in the context of the {\it Gardner transition}~\cite{charbonneau2014fractal, 
charbonneau2015numerical, berthier2016growing, urbani2023gardner}, where  hierarchical structures are found in multiple glass (instead of crystal) states.



{\bf Discussion.}
The proposed TDLM, similar to physical systems such as spin glasses, is important for interpretable deep learning, because it can be investigated using well-developed statistical physics tools. 
The performance advantage or disadvantage of the TDLM, compared to standard DNNs trained by stochastic gradient descent dynamics, shall be examined  in future studies.  
New strategies  inspired by the solid-liquid-solid network structure may improve the training  efficiency of  sophisticated, large-scale models.\\

{\bf Acknowledgments}
We acknowledge financial support from NSFC (Grants 12161141007, 11935002,
and 12047503), from 
Chinese Academy of Sciences (Grant ZDBS-LY-7017), 
and from Wenzhou Institute (Grant WIUCASQD2023009). In this work access was granted to the High-Performance Computing Cluster of Institute of Theoretical Physics - the Chinese Academy of Sciences.\\


\bibliography{bibfile_dnn}

\clearpage


\centerline{\bf \large End Matter}

{\bf Appendix A: Input dataset.}
The input dataset contains the MNIST handwritten digit images~\cite{LeCun, Qiao2007}, which are widely used in machine learning benchmarking.
The original images are coarse-grained into $28\times 28$-pixel pictures, and 
each pixel is either white ($+1$) or black ($-1$). 
We only select images of two digits, ``0'' and ``1'', from the database, for a simple task of binary 
 classification. 
The input dataset is divided into training (2000 images) and test (400 images) parts. \\

\begin{figure}[!htbp]
    \centering
    \includegraphics[width=\linewidth]{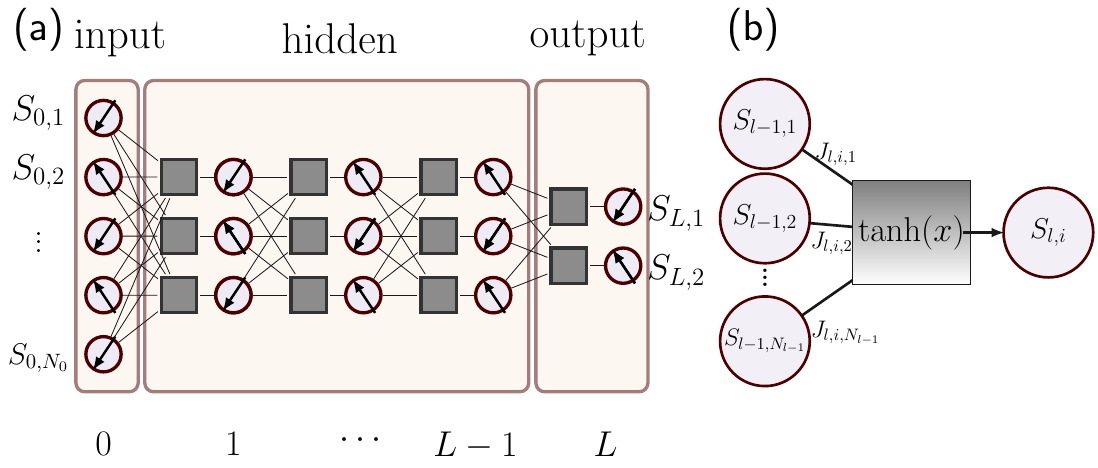} 
    \caption{
    {\bf Illustration of the TDLM.}
    (a) The structure of the fully connected network.
    (b) The basic unit is similar to a single-layer feed-forward perceptron, but $S_{l}$ and  $J_{l}$ are allowed to change dynamically.
    }
    \label{fig:illustration}
\end{figure}

{\bf Appendix B:  thermal deep learning machine.}
The thermal machine is defined on the fully-connected  DNN as shown in Fig.~\ref{fig:illustration}a, where 
$S_{l, i}$ and $J_{l,i,j}$ are real numbers in $(-1, 1)$.
The spins on the boundaries ${\bf S}_0$ and  $ {\bf S}_L$ are fixed by the input/output data respectively, while $ {\bf S}_l$ in other layers ($0<l < L$) and all ${\bf J}_l$ are free variables. The neural network includes three parts: an input layer ($l=0$), multiple hidden layers ($l=1,\cdots, L-1$), and an output layer ($l=L$). 
The input layer, each of the $L-1$ hidden layers, and the output layer consist of $N_0$, $N$, and $N_L$ neurons, respectively. 
Links only exist between neurons in adjacent layers.  
Each neuron in the $l$-th layer is linked to every neuron in the $(l+1)$-th layer. 
Without loss of generality, in this study we set $N_0 = 784$,  $N_L = 2$, $L=10$, and a variable $N$ (unless otherwise specified).

The energy of each layer is defined as,
\begin{equation}
\begin{split}
& \mathcal{E}_l  \left (\{ {\bf S}_{l-1} \},\{ {\bf S}_l \}, \{ {\bf J}_l 
\} \right)  = \\  
 & \sqrt{ \sum_{\mu=1}^M \sum_{i=1}^{N_l} 
\left [S_{l,i}^\mu-\tanh \left (\sum_{j=1}^{N_{l-1}} J_{l,i,j}S_{l-1,j}^\mu \right ) \right]^2}.
\label{eq:H}
\end{split}
\end{equation}
The Hamiltonian of the whole system is,
\begin{equation}
\mathcal{H}  \left (\{ {\bf S}_l \}, \{ {\bf J}_l \} | {\bf S}_0,  {\bf S}_L  \right)  = \sum_{l=1}^{L} \mathcal{E}_l \left( \{ {\bf S}_{l-1} \},\{ {\bf S}_l \}, \{ {\bf J}_l \} \right).
\label{eq:H}
\end{equation}

For a training task of $M$ images, the boundary conditions are fixed by inputs and outputs,
\begin{align}
&{\bf S}_0 \equiv \{S_{0,1}^\mu, S_{0,2}^\mu, \ldots, S_{0,i}^\mu, \ldots, S_{0,N_0}^\mu \}, \\
& {\bf S}_L \equiv \{S_{L,1}^\mu, S_{L, 2}^\mu, \ldots,  S_{L,i}^\mu, \ldots, S_{L,N_L}^\mu \}.
\end{align}
where $\mu = 1,  \ldots, M$. The free (dynamical) variables are the spin and weight parameters,
\beq
{\bf S}_l \equiv \{S_{l,1}^\mu, S_{l,2}^\mu, \ldots, S_{l,i}^\mu, \ldots, S_{l,N_l}^\mu \},
\eeq
 and
\beq
{\bf J}_l \equiv \{J_{l,1,1},  \ldots, J_{l,i,j}, \ldots, J_{l, N_l, N_{l-1}} \}, 
\eeq
where $l = 1, \ldots, L$ and $\mu = 1,  \ldots, M$.
The spin and weight parameters have $\sim LMN$ and $\sim LN^2$ total variables respectively. 

The $\mu=1,\ldots, M$ training images impose $M$ boundary conditions, or constraints, to the problem. We require that $\{ {\bf J}_{l} \}$ are independent of the constraint $\mu$, while $\{ {\bf S}_{l} \}$ can depend on $\mu$. 
Here, by setting $\{ {\bf S}_{l} \}$ as free variables, we allow  the constraints on some perceptrons unsatisfied, at the cost of a non-zero energy. . 
When the total potential energy is zero,  all constraints are satisfied, and the output matches the image label perfectly. 

The corresponding statistical mechanical ensemble is characterized by a partition function, 
\beq
Z  \left( {\bf S}_0 , {\bf S}_L  \right) =  \sum_{\{ {\bf S}_l \}, \{ {\bf J}_l \} } e ^{-\beta \mathcal{H} \left (\{ {\bf S}_l \}, \{ {\bf J}_l \} |  {\bf S}_0 ,  {\bf S}_L  \right )},
\label{eq:patition}
\eeq
where $\beta = 1/(k_{\rm B} T)$ is the inverse temperature. 
From Eq.~(\ref{eq:patition}), one can see that the boundary  conditions 
$\left(  {\bf S}_0,  {\bf S}_L \right)$  play the role of quenched disorder in this model.\\

{\bf Appendix C: molecular dynamics training algorithm.}
To train the thermal machine, we perform classical MD simulations following Langevin dynamics, at a fixed low temperature $k_B T=10^{-5}$ (unless otherwise specified), where  the Boltzmann constant $k_{\rm B}=1$. 
The spin and weight variables $\{ {\bf S}_{l} \}$ and $\{ {\bf J}_{l} \}$ are then considered as positions $\{ r_i \}$ of atoms, updated by the formulas,
\begin{equation}
    v_i(t+\Delta t/2)=v_i(t-\Delta t/2)+F_i(t) \frac{\Delta t}{m_i},
\end{equation}
and
\begin{equation}
    r_i(t+\Delta t)=r_i(t)+v_i(t+\Delta t/2)\Delta t,
\end{equation}
where $v_i$ is the velocity, the mass $m_i=1$ for all $i$, and the time step $\Delta t=10^{-2}$.
The force is given by,
\begin{equation}
    F_i(t)=-\frac{dE(r_1, r_2, \ldots; t)}{dr_i}-\eta v_i(t-\Delta t/2)+ \sqrt{\eta(2-\eta)k_B T}\delta_i,
\end{equation}
where $E(r_1, r_2, \ldots; t)$ is the total energy computed according to Eq.~(\ref{eq:H}) at time $t$,  $\eta=10^{-3}$ is the friction coefficient,  and $\delta_i$ is a random variable with the standard normal distribution with a zero mean and a unit standard deviation.  Because spin and weight variables are restricted in the range $(-1, 1)$,  we apply reflective boundary conditions (elastic hard walls) to the simulation box.
The program is accelerated using the GPU-based PyTorch software, which performs differentiation (MD force calculations) automatically and efficiently~\cite{paszke2017automatic}.

The initial conditions at $t=0$ are random configurations of  $\{ {\bf S}_{l} \}$ and $\{ {\bf J}_{l} \}$ drawn from uniform distributions in $(-1, 1)$. Each independent initial condition is called a { sample}. The dynamical data presented in this work are averaged over $N_{\rm s}$ samples. For a given sample, $N_{\rm r}$ {replicas} are created at a waiting time $t_{\rm w}$. These replicas share the same particle positions at  $t_{\rm w}$, but are assigned different velocities at  $t_{\rm w}$ drawn from the Maxwell-Boltzmann distribution, thus following different dynamical trajectories after $t_{\rm w}$.\\

{\bf Appendix D: Feed-forward neural networks for testing.} 
We test the $\{ {\bf J}_l \}$ configurations learned by the TDLM in a standard feed-forward DNN.
The basic unit is a single-layer perceptron,
\beq
S_{l, i} = \text{sgn}\left(\frac{1}{\sqrt{N_{l-1}}}\sum_{j=1}^{N_{l-1}} J_{l, i, j} S_{l-1, j}\right),
\label{eq:activation0}
\eeq
where $N_{l-1}$ is the number of neurons in the $(l-1)$-th layer.
 Each neuron is either active ($S_{l, i} = 1$) or passive ($S_{l, i}  = -1$), and the weights 
 $J_{l,i,j} \in (-1, 1)$ are copied from the TDLM. Note that $\{ {\bf J}_{l}(t) \}$ evolves with time during the MD training, which allows us to test the time-dependent test accuracy $A(t)$ by copying the instantaneous  $\{ {\bf J}_{l}(t) \}$ to the test DNN. \\

{\bf Appendix E: Definition of time correlation functions.} We define two types of time correlation functions. The first type is the   {\it auto-correlation functions} for $ {\bf S}_l $ and $ {\bf J}_l $, defined as,
\begin{align}
c_l(t,t_{\rm w}) &= \frac{1}{N_l M}\sum_{i=1}^{N_l}\sum_{\mu=1}^{M}  S_{l,i}^\mu(t+t_{\rm w}) S_{l,i}^\mu(t_{\rm w}),\\
C_l(t,t_{\rm w}) &= \frac{1}{N_l N_{l-1}}\sum_{i=1}^{N_l}\sum_{j=1}^{N_{l-1}}  J_{l,i,j}(t+t_{\rm w}) J_{l,i,j} (t_{\rm w}), 
\end{align}
where $t_{\rm w}$ is the waiting time, $t$  the time collapsed after waiting, and $t+t_{\rm w}$ the total time. Similarly, the global  auto-correlation functions are,
\begin{align}
c(t,t_{\rm w}) &= \frac{1}{L-1} \sum_{l=1}^{L-1}\left[\frac{1}{N_l M}\sum_{i=1}^{N_l}\sum_{\mu=1}^{M}  S_{l,i}^\mu(t+t_{\rm w}) S_{l,i}^\mu(t_{\rm w}) \right],\\
C(t,t_{\rm w}) &= \frac{1}{L} \sum_{l=1}^{L} \left[ \frac{1}{N_l N_{l-1}}\sum_{i=1}^{N_l}\sum_{j=1}^{N_{l-1}}  J_{l,i,j}(t+t_{\rm w}) J_{l,i,j} (t_{\rm w}) \right].
\end{align}
The auto-correlation functions characterize how fast the system relaxes. 


The second type is the {\it replica-correlation functions} defined as,
\begin{align}
q_l^{ab}(t,t_{\rm w}) &= \frac{1}{N_l M}\sum_{i=1}^{N_l}\sum_{\mu=1}^{M}  S_{l,i}^{\mu,a}(t+t_{\rm w}) S_{l,i}^{\mu,b}(t+t_{\rm w}),
\label{eq:q}\\
Q_l^{ab}(t,t_{\rm w}) &= \frac{1}{N_l N_{l-1}}\sum_{i=1}^{N_l}\sum_{j=1}^{N_{l-1}} J_{l,i,j}^{a}(t+t_{\rm w}) J_{l,i,j}^{b}(t+t_{\rm w}), 
\label{eq:Q}
\end{align}
where $a, b = 1, 2, \ldots, N_{\rm s} \times N_{\rm r}$ are  pairs of replicas, which are generated at $t_{\rm w}$ and evolve differently after that. In this study,  $N_{\rm s} = 8-40$, and $N_{\rm r} = 6-40$. The $q_l(t,t_{\rm w})$ and $Q_l(t,t_{\rm w})$  are $q_l^{ab}(t,t_{\rm w})$ and $Q_l^{ab}(t,t_{\rm w})$ averaged over $ab$ pairs of replicas.
Correspondingly, the global  replica-correlation functions are,
\begin{align}
q^{ab}(t,t_{\rm w}) &= \frac{1}{L-1} \sum_{l=1}^{L-1}\left[\frac{1}{N_l M}\sum_{i=1}^{N_l}\sum_{\mu=1}^{M}  S_{l,i}^{\mu,a}(t+t_{\rm w}) S_{l,i}^{\mu,b}(t+t_{\rm w}) \right],\\
Q^{ab}(t,t_{\rm w}) &= \frac{1}{L} \sum_{l=1}^{L} \left[ \frac{1}{N_l N_{l-1}}\sum_{i=1}^{N_l}\sum_{j=1}^{N_{l-1}} J_{l,i,j}^{a}(t+t_{\rm w}) J_{l,i,j}^{b}(t+t_{\rm w}) \right].
\end{align}

{\bf Appendix F: Clustering algorithm to construct the overlap matrix and the corresponding tree.}
We employ the Complete Linkage (CL) clustering  algorithm~\cite{Hastie2009} to generate the heat map of the overlap parameter matrix and the corresponding tree in Fig~\ref{fig:overlap_heatmap_dendrogram}. The method groups data points into hierarchical clusters based on their maximum pairwise distances. We define the distance between every pair of replicas in the $l$-th layer, using a metric $d^{ab}_l \equiv 1-Q^{ab}_l$. 
The CL intercluster distance (dissimilarity) is the largest distance between any member $a$ from the first cluster $A$ (i.e., $a \in A$) and any member $b$ from the second cluster $B$ (i.e., $b \in B$), $D_l^{AB} = \max_{a \in A, b \in B} d_l^{ab}.$

The clustering process is outlined as follows.

{\it i) Generating initial clusters.}
Consider each replica as an individual cluster. 

{\it ii) Finding the closest clusters.}
Find a pair of clusters $A$ and $B$ with the minimum CL intercluster distance $D_l^{AB}$.

{\it iii) Merging clusters.}
Merge the two closest clusters $A$ and $B$ into $C$. Update the CL intercluster distances between $C$ and other clusters. Continue merging the closest clusters until all data points are consolidated into a single cluster.

{\it iv) Constructing a dendrogram.}
Construct a dendrogram that records the sequence of merging and the corresponding CL intercluster distances. This sequence can be visualized by a tree as shown in  Fig~\ref{fig:overlap_heatmap_dendrogram}.

The above algorithm is implemented by using the \texttt{dendrogram} and \texttt{linkage} functions in the \texttt{scipy.cluster.hierarchy} module of the SciPy program~\cite{virtanen2020scipy}.


\end{document}



\title{Liquid and solid layers in a thermal deep learning machine - Supplementary Material}


\author{Gang Huang}
\affiliation{Department of Physics, Chengdu University of Technology, Chengdu 610059, China}
\affiliation{Institute of Theoretical Physics, Chinese Academy of Sciences, Beijing 100190, China}

\author{Lai Shun Chan}
\affiliation{Department of Physics, City University of Hong Kong, Hong Kong Special Administrative Region of China, People's Republic of China}

\author{Hajime Yoshino}
\affiliation{D3 Center, Osaka University, Toyonaka, Osaka 560-0043, Japan}
\affiliation{Graduate School of Science, Osaka University, Toyonaka, Osaka 560-0043, Japan}

\author{Ge Zhang}
\email{gzhang37@cityu.edu.hk}
\affiliation{Department of Physics, City University of Hong Kong, Hong Kong Special Administrative Region of China, People's Republic of China}

\author{Yuliang Jin}
\email{yuliangjin@mail.itp.ac.cn}
\affiliation{Institute of Theoretical Physics, Chinese Academy of Sciences, Beijing 100190, China}
\affiliation{School of Physical Sciences, University of Chinese Academy of Sciences, Beijing 100049, China}
\affiliation{Wenzhou Institute, University of Chinese Academy of Sciences, Wenzhou, Zhejiang 325000, China}

\date{\today}

\begin{abstract}
\end{abstract}

\maketitle

\tableofcontents
\setcounter{figure}{0}
\setcounter{equation}{0}
\setcounter{table}{0}
\renewcommand\thefigure{S\arabic{figure}}
\renewcommand\theequation{S\arabic{equation}}
\renewcommand\thesection{S\arabic{section}}
\renewcommand\thetable{S\arabic{table}}

\clearpage

\section{Dependence of the training energy and test accuracy on the temperature}

\begin{figure*}[!htbp]
  \centering
\includegraphics[width= 0.8\linewidth]{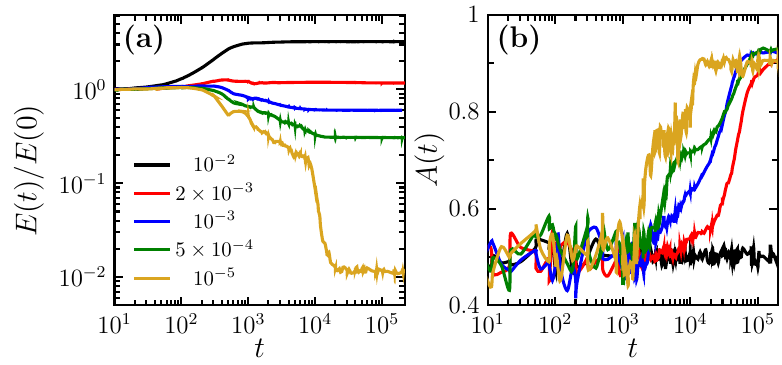}
  \caption{{\bf Training energy  $E(t)/E(0)$ and test accuracy $A(t)$ at a few different  temperatures $T$.} We set $M=2000$ and $\ln \alpha = 4.6$.}
 \label{fig:T_dependence_lna4}
\end{figure*}

Figure~\ref{fig:T_dependence_lna4} compares the training  energy (loss function) $E(t)/E(0)$ and test accuracy $A(t)$ at  different  temperatures $T$. In all cases, the initial configurations at $t=0$ are random ($T=\infty$), and the system is rapidly quenched to the target temperature $T$. 
Except for the highest $T=10^{-2}$, the asymptotic test accuracy at the large time limit is nearly independent of $T$ (Fig.~\ref{fig:T_dependence_lna4}b).

\clearpage

\section{Additional results on global correlation functions}
\begin{figure*}[!htbp]
  \centering
 \includegraphics[width= \linewidth]{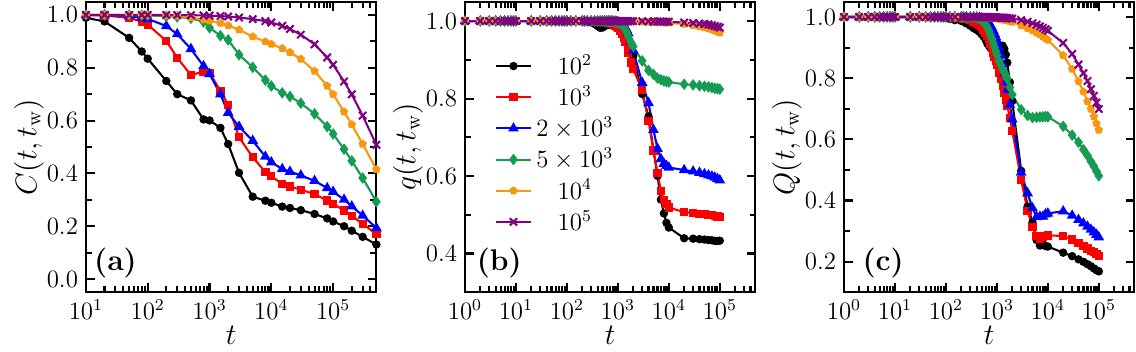}
  \caption{{\bf Global correlation functions.}
  (a) Auto-correlation function $C(t,t_{\rm w})$ of weights. (b) Replica-correlation function $q(t,t_{\rm w})$ of spins. (c) Replica-correlation function $Q(t,t_{\rm w})$ of weights. The legend indicates $t_{\rm w}$. We set  $M=2000$ and $\ln \alpha = 5.3$.
  }
 \label{fig:other_global_correlation}
\end{figure*}

In Fig.~2g we have shown the global auto-correlation function $c(t,t_{\rm w})$ of spins. The results for the other three types of global correlation functions,  $C(t,t_{\rm w})$, $q(t,t_{\rm w})$ and $Q(t,t_{\rm w})$, are presented in Fig.~\ref{fig:other_global_correlation}. 

\clearpage

\section{Additional results on layer-specific correlation functions}
\begin{figure*}[!htbp]
  \includegraphics[width= \linewidth]{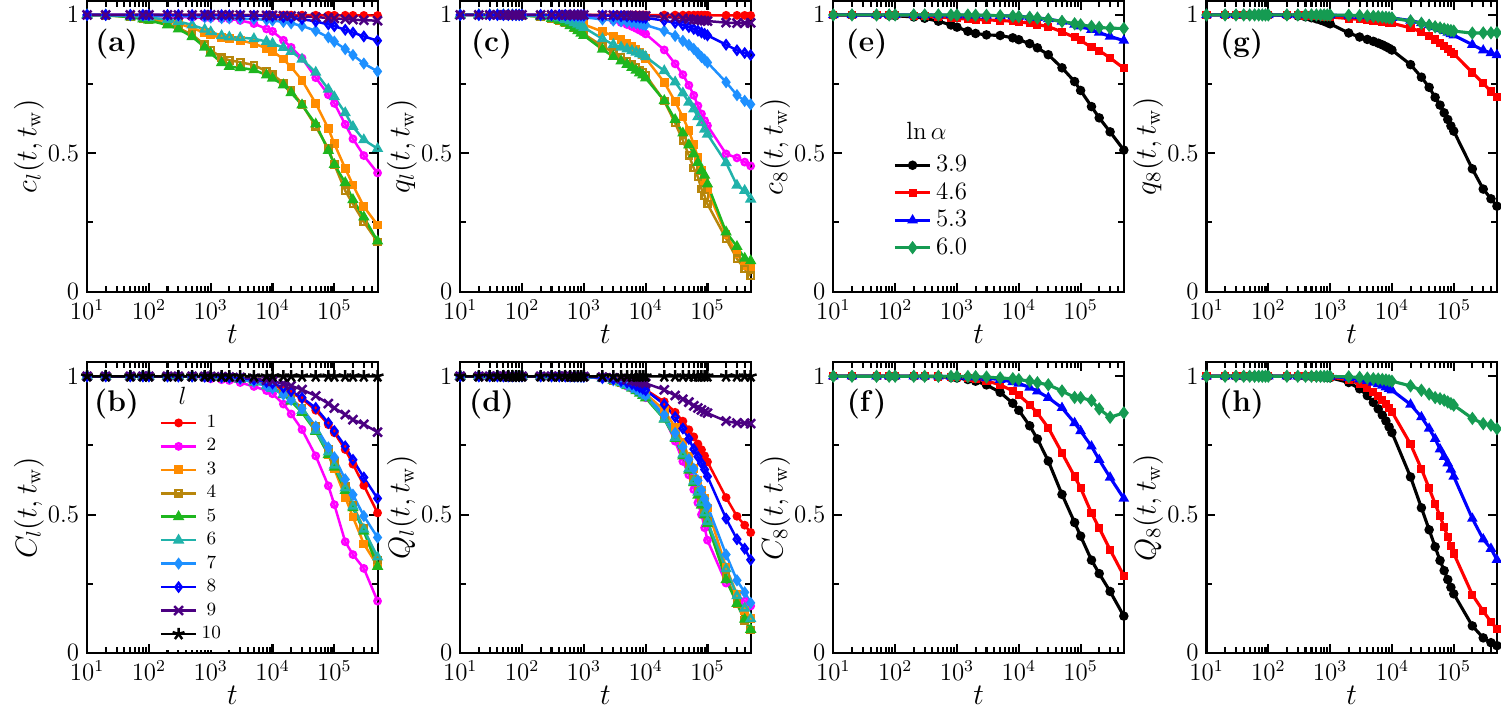}
  \caption{
  {\bf Layer-specific correlation functions.}
  Data are obtained with a fixed waiting time $t_{\rm w}=5 \times 10^4$.
  (a) Spin auto-correlation functions $c_l(t, t_{\rm w})$, (b) weight auto-correlation functions $C_l(t, t_{\rm w})$, (c) spin replica-correlation functions $q_l(t, t_{\rm w})$, and (d) weight replica-correlation functions $Q_l(t, t_{\rm w})$, for different $l$, with a fixed $\ln\alpha=5.3$. 
  (e-h). Correlation functions for the $8$-th layer, with a few different $\ln \alpha$. 
 }
 \label{fig:layer_additional}
\end{figure*}

\begin{figure*}[!htbp]
   \centering
   \includegraphics[width= 0.5\linewidth]{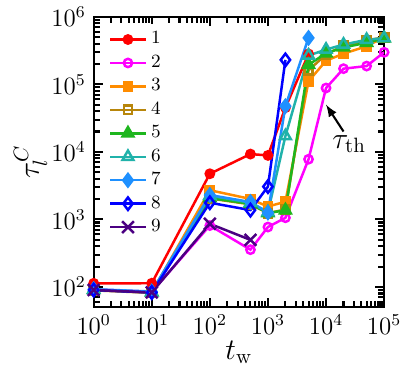}
   \caption{{\bf Waiting time $t_{\rm w}$-dependence of the weight relaxation time $\tau^C_l$.} The legend indicates $l$. We set $\ln \alpha = 5.3$.
   }
   \label{fig:tau_tw_lna5}
\end{figure*}

In Fig.~\ref{fig:layer_additional}, we present additional data on the layer-specific correlation functions. 
The waiting time $t_{\rm w}$-dependence of the weight relaxation time $\tau_l^C$ is shown in Fig.~\ref{fig:tau_tw_lna5}, where $\tau_l^C$ is defined by $C_l(t=\tau_l^C, t_{\rm w}) = 1/e$ (see Fig.~3b for corresponding data on  $\tau_l^c$). The data, together with Fig.~3, consistently suggest a solid-liquid-solid structure of parameter states in the network.

\begin{figure*}[!htbp]
  \centering
  \includegraphics[width= 0.9\linewidth]{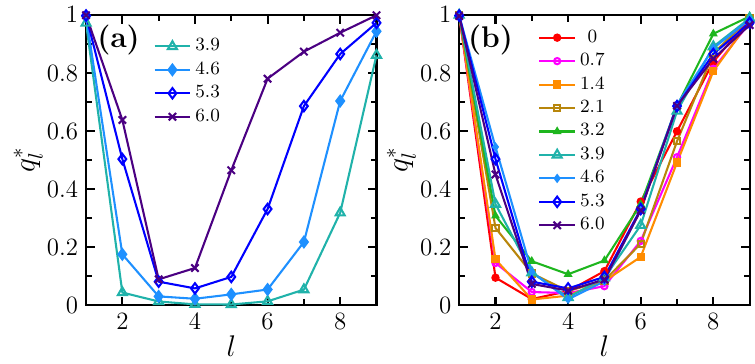}
  \caption{{\bf Solid-liquid-solid network structure for different storage ratio $\alpha$.}
   We plot $q_l^*\equiv q_l(t=5 \times 10^5, t_{\rm w}=5 \times 10^4)$ as a function of $l$, for a few different $\ln \alpha = \ln (M/N)$ as indicated in the legend.
  (a) $M=2000$ fixed and $N$ varied. (b) $N=10$ fixed and $M$ varied. 
  } 
 \label{fig:finite_size}
\end{figure*}

When $\alpha$ increases, the correlation functions decay slower, but the solid-liquid-solid structure is robustly observed (Fig.~\ref{fig:finite_size}). In Fig.~\ref{fig:finite_size}a, in order to vary $\alpha = M/N$, we fix $M=2000$ and adjust correspondingly $N$ (see Table~\ref{table:m_n_given_alpha}). It can be shown that varying $M$ with a fixed $N=10$ creates a less significant change to the network behavior (Fig.~\ref{fig:finite_size}b). These results reveal important differences between random and MNIST data. In the former, only the ratio $\alpha = M/N$ is relevant~\cite{Yoshino2020}. In contrast, for structured data such as MNIST images, the number of constraints is effectively smaller than $M$, $M_{\rm eff} < M$, due to significant data correlations. Thus, for a fixed $N$, varying $M$ does not introduce a big difference when  $M_{\rm eff} < M$ (Fig.~\ref{fig:finite_size}b). On the other hand, changing $N$ can introduce noticeable finite-size effects as observed in Ref.~\cite{yoshino2023spatially}.

\begin{table*}[!htbp]
\centering
\caption{{\bf Parameters of $M$, $N$, and $\ln \alpha = \ln (M/N)$ used in this study.} Two settings are listed: fixed $M=2000$ and fixed $N=10$. For the fixed $M=2000$, the maximum $N$ can be simulated by the current algorithm is $N_{\rm max}=40$.}
\begin{tabular}{l  l    l   l  l}
\hline
\hline
 $\ln\alpha$  & $M$ (fixed)  & $N$ & $M$ & $N$ (fixed) \\
\hline
 0 & - & - & 10 & 10 \\
 0.7 & - & - & 20 & 10 \\
 1.4 & -  & - & 40 & 10 \\
 2.1 & -  & - &  80 & 10\\
 2.8 & -  & - & 160 & 10\\
 3.2 & - & - & 250 & 10\\
 3.9 & 2000 & 40 & 500 & 10\\
 4.6 & 2000 & 20 & 1000 & 10\\
 5.3 & 2000 & 10 & 2000 & 10\\
 6.0 & 2000 & 5 & 4000 & 10\\
\hline 
\hline  
\end{tabular}
\label{table:m_n_given_alpha}
\end{table*}

\clearpage

\section{Dependence on the total network depth}
\begin{figure*}[!htbp]
  \centering
\includegraphics[width= \linewidth]{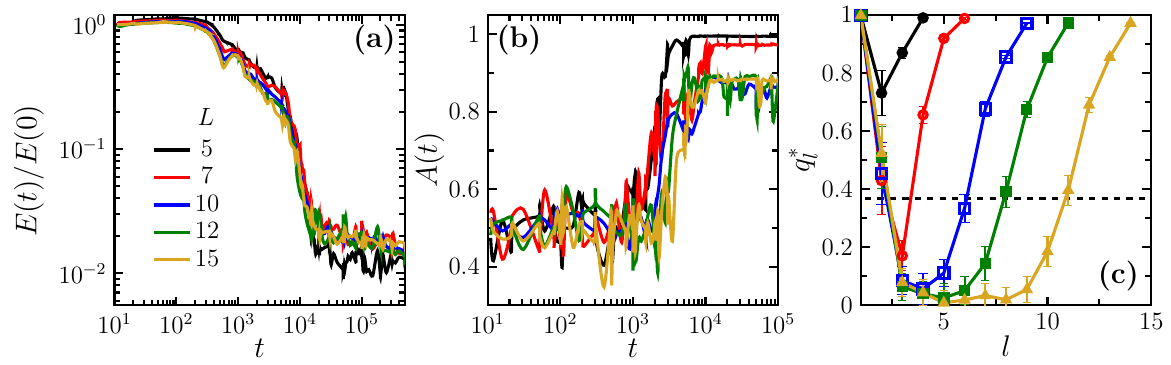}
  \caption{{\bf Effects of the total network depth $L$.}
  (a) Training energy density $E(t)/E(0)$ and (b) test accuracy $A(t)$ for different $L$ ($\ln \alpha = 5.3$). (c) Spin replica-correlation $q_l^{*}$ = $q_l(t = 5 \times 10^5,t_{\rm w} = 5 \times 10^4)$ as a function of $l$, for different $L$.
    The dashed horizontal line corresponds to $1/e$.
  }
 \label{fig:L_dependence}
\end{figure*}

In Fig.~\ref{fig:L_dependence}, the effects of the total network depth $L$ are examined. Interestingly, we observe that when the network is deep enough, i.e., when $L \geq 10$ (for $\ln \alpha =5.3$), the depth $\xi_{\rm s}$ of solid layer does not change anymore even $L$ is further increased. In such deep networks, increasing $L$ only introduces more liquid layers. 

\clearpage

\section{Additional data on $P(Q_l^{ab})$}
\begin{figure*}[!htbp]
  \centering
\includegraphics[width= \linewidth]{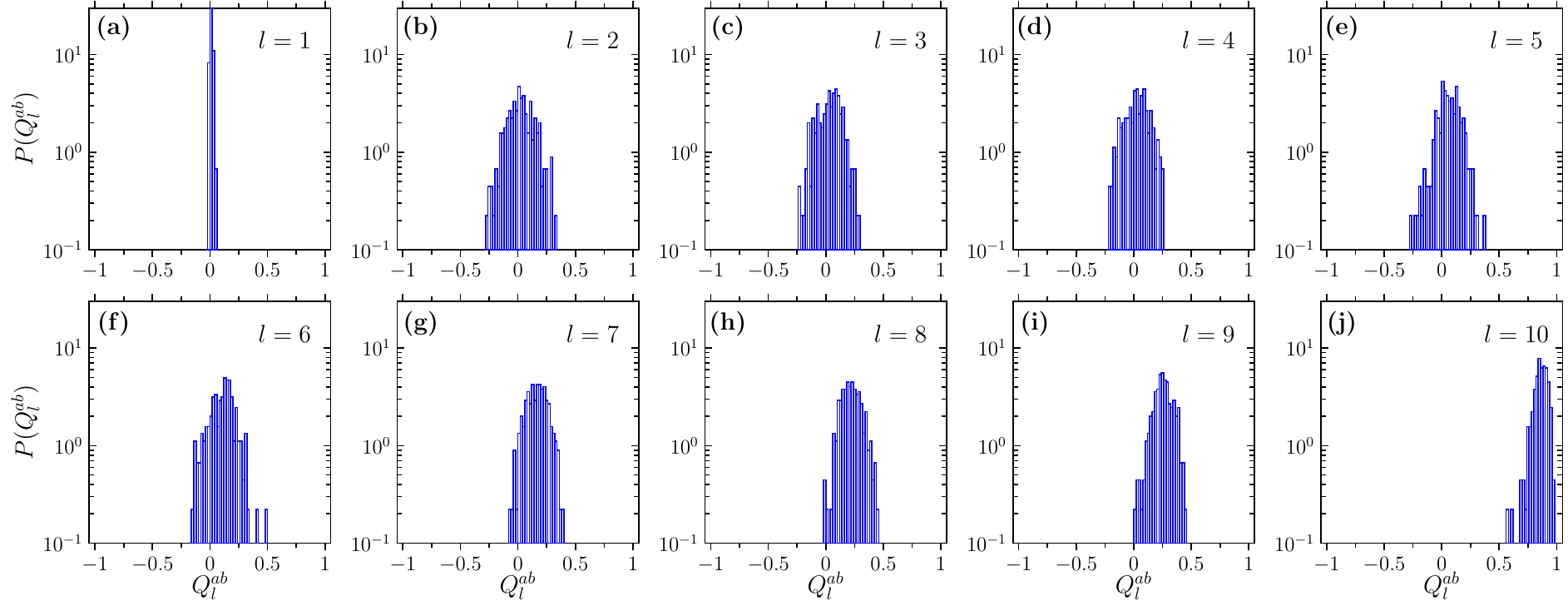}
  \caption{{\bf $P(Q_l^{ab})$ in different layers.}
  We define $Q_l^{ab} \equiv Q_l^{ab}(t=5\times 10^5, t_{\rm w}=10^3)$, and set $\ln \alpha = 0$.
  }
 \label{fig:additional_PQl_0}
\end{figure*}

\begin{figure*}[!htbp]
  \centering
\includegraphics[width= \linewidth]{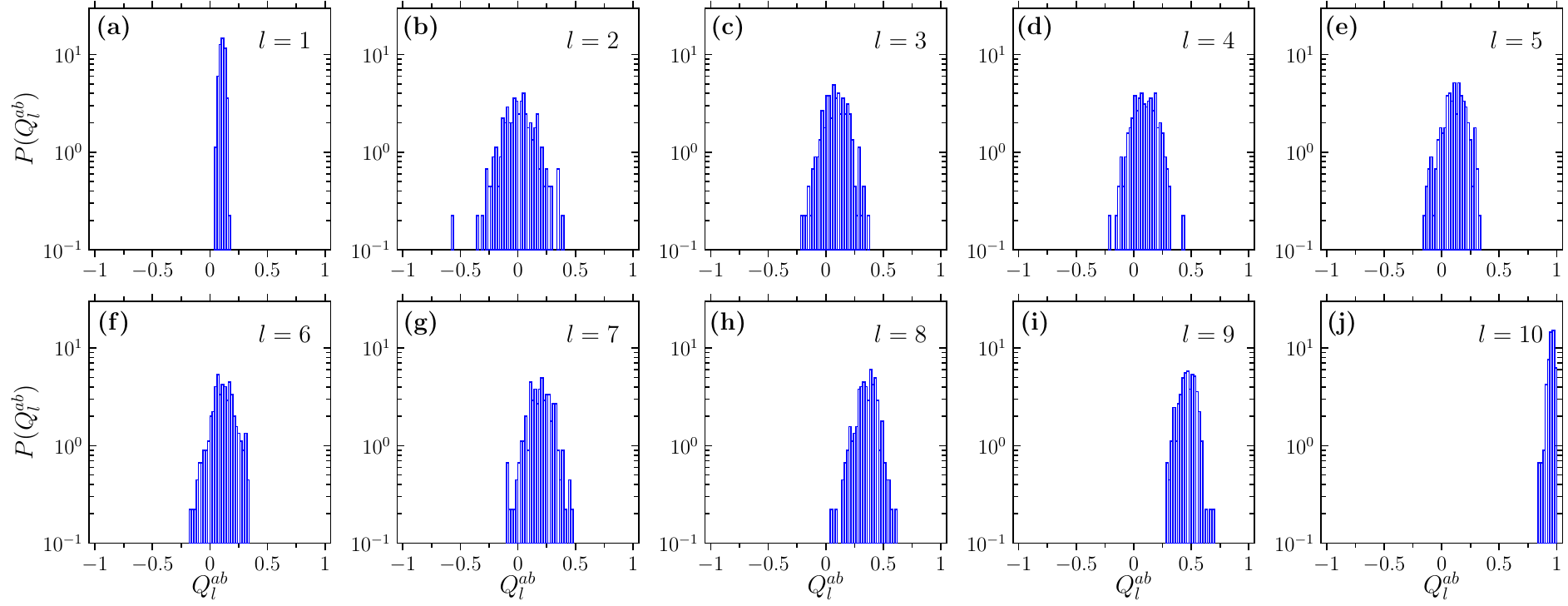}
  \caption{{\bf $P(Q_l^{ab})$ in different layers.}
  We define $Q_l^{ab} \equiv Q_l^{ab}(t=5\times 10^5, t_{\rm w}=10^3)$, and set $\ln \alpha = 3.2$.
  }
 \label{fig:additional_PQl_3.2}
\end{figure*}

\begin{figure*}[!htbp]
  \centering
\includegraphics[width= \linewidth]{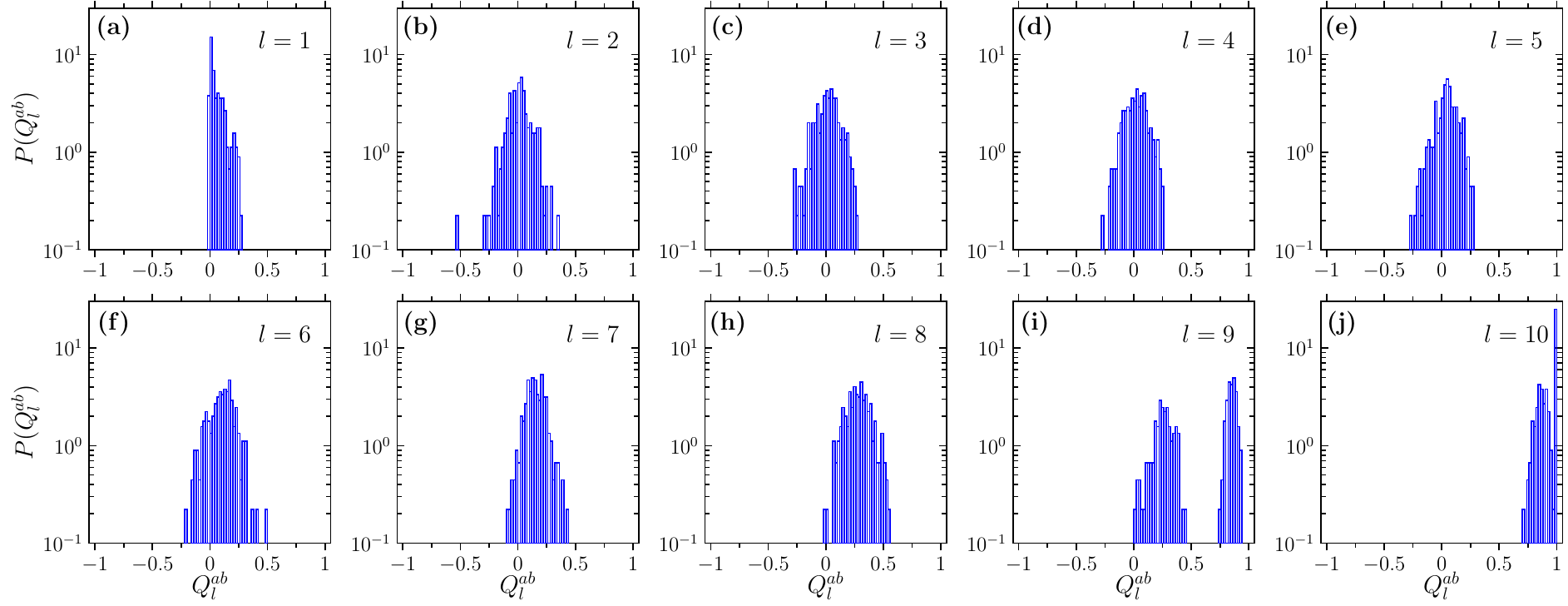}
  \caption{{\bf $P(Q_l^{ab})$ in different layers.}
  We define $Q_l^{ab} \equiv Q_l^{ab}(t=5\times 10^5, t_{\rm w}=10^3)$, and set $\ln \alpha = 6.0$.
  }
 \label{fig:additional_PQl_6.0}
\end{figure*}

In Figs.~\ref{fig:additional_PQl_0},~\ref{fig:additional_PQl_3.2} and~\ref{fig:additional_PQl_6.0}, we plot the distributions $P(Q_l^{ab})$  in different layers ($l=1-10$), for $\ln \alpha = 0, 3.2, 6.0$ (for a fixed $N=10$). 
(i) When $\ln \alpha \leq 3.2$, $P(Q_l^{ab})$ is close to a Gaussian distribution in all layers. The center of the Gaussian distribution moves with $l$, which behaves similarly as the $q_l^*$ data in Fig.~3d.
(ii) When $\ln \alpha > 3.2$, only the liquid layers still have a Gaussian-like $P(Q_l^{ab})$.
Approaching the input or output boundary, a non-trivial distribution gradually emerges, suggesting a hierarchical evolution of the design space from a simple structure in liquid layers to a complex structure in solid layers~\cite{Yoshino2020}. 

\clearpage

\bibliography{bibfile_si}